\documentclass[aps,pre,twocolumn,twoside, amsmath,amssymb,aps,bbm,floatfix,showpacs]{revtex4-1}

\usepackage{xcolor}
\usepackage{graphicx}
\usepackage{subfigure}
\usepackage{dcolumn}
\usepackage{bm}
\usepackage{framed}
\usepackage{hyperref}
\usepackage{dsfont}
\usepackage[normalem]{ulem}
\usepackage{nicefrac}

\usepackage{nicefrac}
\usepackage{braket}

\newcommand{\be}{\begin{equation}}
\newcommand{\ee}{\end{equation}}
\newcommand{\bea}{\begin{eqnarray}}
\newcommand{\eea}{\end{eqnarray}}

\begin{document}

\title{Power fluctuations in  a finite-time quantum Carnot engine}
\author{Tobias Denzler}
\author{Eric Lutz}
 \affiliation{Institute for Theoretical Physics I, University of Stuttgart, D-70550 Stuttgart, Germany}

\begin{abstract}
Stability is an important property of small thermal machines with fluctuating power output. We here consider a finite-time quantum Carnot engine based on a degenerate multilevel system and study the influence of its finite Hilbert space structure on its stability. We  optimize in particular its relative work fluctuations with respect to level degeneracy and level number. We find that its optimal performance may surpass those of  nondegenerate two-level engines or harmonic oscillator motors. Our results show how to realize high-performance, high-stability cyclic quantum heat engines.
\end{abstract}

\maketitle

The Carnot engine is one  of the most emblematic examples of a thermal machine. Since its introduction  in 1824, it has become  a representative  model for all heat engines. The Carnot cycle  simply consists of two isothermal (expansion and compression) steps and of two adiabatic (expansion and compression) processes \cite{cen01}. In its ideal reversible limit of infinite cycle duration, the Carnot motor is the most efficient  heat engine. The corresponding Carnot efficiency, $\eta_\text{C} = 1-T_c/T_h$, where $T_{c,h}$ are the respective temperatures of the cold and hot heat baths, is regarded as the first formulation of the second law of thermodynamics \cite{cen01}. The first experimental realization of a Carnot engine using a colloidal particle  in an optical harmonic trap  has been reported lately \cite{mar15}. In addition, the finite-time properties of the Carnot cycle have been well investigated theoretically both in the classical \cite{cur75,and77,gut78,rub79,sal81,ond83,esp10,cav17} and in the  quantum \cite{gev92,gev92a,wu06,qua07,abe11,wan12,all13,dan20,abi20} regimes. Strong  emphasis has been put on the optimization of the performance of the   engine for finite cycle durations, in particular on its average power output at the expense of its efficiency.

For classical microscopic  heat engines, like the one implemented in the experiment of Ref.~\cite{mar15}, thermal fluctuations are no longer negligible  as is the case for  macroscopic motors \cite{sei12}. As a result, key performance measures, such as efficiency \cite{ver14,ver14a,pol15,man19} and power \cite{hol14,hol17,pie18,hol18}, are stochastic variables. In that context, attention has recently been  drawn to power fluctuations as a limiting factor for the practical usefulness of thermal machines: heat engines should indeed ideally have high efficiency, large power output but small power fluctuations \cite{hol14,hol17,pie18,hol18}. A new figure of merit,  the constancy, defined as the product of the variance of the power and time, has been introduced to characterize the stability of heat engines \cite{hol14,hol17,pie18,hol18}. While a strict trade-off between efficiency, power and constancy has been established for steady-state heat engines, implying that power fluctuations diverge at maximum efficiency and finite power \cite{pie18}, they may remain finite for quasistatic cyclic thermal machines \cite{hol18}. On the other hand, quantum motors are not only dominated by thermal fluctuations but also by quantum fluctuations. 

In this paper, we investigate the generic features of the power fluctuations in a finite-time quantum Carnot engine in the quasistatic limit. We specifically study the interplay between power  fluctuations, the finite dimensionality of the Hilbert space of the working medium and the degree of degeneracy of its levels. Degenerate finite level structures commonly appear in atomic \cite{foo05}, molecular \cite{bro13} and condensed-matter physics \cite{cha00}. An understanding of their influence on the stability of quantum heat engines is therefore essential for future experimental realizations of thermal devices in these systems \cite{baz03}. An important illustration of the effect of the finiteness of  quantum systems on thermodynamic fluctuations is provided by the Schottky anomaly \cite{blu06}: the corresponding increase of the heat capacity at low temperatures does not occur in infinite dimensional systems like the harmonic oscillator, in which the energy is not bounded; it is, furthermore, strongly affected by  level degeneracy and level number \cite{eva06}. We mention, however,  that our results are not directly related to the Schottky anomaly. 

In the following, we compute the inverse coefficient of variation for work, defined as the ratio of the mean work and its standard deviation \cite{eve98}, for a quasistatic quantum Carnot engine whose working medium is described by a homogeneous Hamiltonian of degree $-2$, ${\cal H}(b \bm{r}) = b^{-2}{\cal H}(\bm{r})$.  Such Hamiltonians characterize  a large class   of single-particle, many-body and nonlinear systems that exhibit equidistant spectra \cite{gri10,jar13,cam13,def14,bea16}.
We obtain a general formula that only depends on  the heat capacity and on the entropy variation during the hot isotherm. We use this expression to maximize the inverse coefficient of variation for work, with respect to the degree of degeneracy and the number of levels of the system, in order to attain optimal cyclic quantum  engines that operate close to  the Carnot efficiency with large power output and small power fluctuations. We illustrate our results by  analyzing  (i) a two-level system with arbitrary degeneracy, (ii) a nondegenerate system with arbitrary level number, and (iii) a three-level system with generic degree of degeneracy. In all  cases the optimal inverse coefficient of variation for work may be numerically determined by  solving a  transcendental equation.

\textit{Finite-time quantum Carnot cycle.}
We consider a general quantum system with time-dependent Hamiltonian, $H_t=\omega_t {\cal H}$, as the working fluid of a finite-time quantum Carnot engine, with driving parameter $\omega_t$. 
The quantum Carnot cycle consists of the following four steps \cite{gev92,gev92a,wu06,qua07,abe11,wan12,all13,dan20,abi20}: (1) hot isothermal expansion from $\omega_1$ to $\omega_2$, at  temperature $T_h$  in time $\tau_1$, during which  work $W_1$ is produced by the system and heat $Q_h$ is absorbed; (2)  adiabatic expansion from $\omega_2$ to $\omega_3$, in time $\tau_2$, during which work $W_2$ is  performed and the entropy remains constant. The system is here decoupled from the baths and its Hamiltonian commutes with itself at all times, $\left[H_t,H_{t'}\right]=0$. As a result, nonadiabatic transitions do not occur for all driving times $\tau_2$; (3) cold isothermal compression  from $\omega_3$ to $\omega_4$, at  temperature $T_c$  in time $\tau_3$, during which  work $W_3$ is done on the system and heat $Q_c$ is released; (4) adiabatic compression from $\omega_4$ to $\omega_1$, in time $\tau_4$, during which work $W_4$ is  unitarily performed on the system. The total cycle time is $\tau= \sum_i \tau_i$. Work and heat are taken positive when added to the system. 

In order to evaluate the finite-time performance of the  quantum Carnot engine, we will  determine the mean and the variance of the stochastic  work output $w$. The average  total work, $W = \langle w \rangle = \sum_i W_i$, directly follows from the combination of the first and second law   \cite{cen01},
  \begin{equation}
  \label{1}
  	W  = -Q_h-Q_c=(T_c - T_h)\Delta S,
  \end{equation}
 where $\Delta S$ denotes the entropy change   during the hot isotherm. Meanwhile, the total work fluctuations are characterized by the probability distribution  $P(w)$,
  \begin{equation}\label{eq:pw1}
  	P(w) = \langle \delta (w- \sum_i w_i ) \rangle,
  \end{equation}
 where the average is taken over  the joint probability distribution  given by the convolution of the work densities  of the four branches of the cycle, $P(w_1,w_2,w_3,w_4) =  P_1(w_1) \star P_2(w_2) \star P_3(w_3) \star P_4(w_4)$ \cite{hol18}. The two  isotherms (1) and (3) are assumed to be slower than the (fast) relaxation induced by the    baths. The system thus remains in a thermal state and the two finite-time processes are quasistatic. In this case, the work distributions are sharp (with no fluctuations) and work is deterministic \cite{sup},
   \begin{equation}
    P_{1,3}(w_{1,3})= \delta\left(w_{1,3}-W_{1,3}\right).
  \end{equation}
 On the other hand, since no heat is exchanged during the two unitary adiabats (2) and (4), the corresponding work distributions can be obtained via the usual  two-point-measurement scheme by projectively measuring the energy   at the beginning and at the end of each step \cite{tal07}.  Without level transitions, we obtain for process (2),
\begin{equation}
	P_2(w_2)=\sum_{n} \delta \left[w_2 - (E_3^{n}-E_{2}^{n})\right] p^{n}_2,
\end{equation}  
where $E_{2}^n$  and $E_{3}^n$ denote the respective eigenvalues of the Hamiltonians $H_2 = H_{\tau_1}$ and $H_3= H_{\tau_1+ \tau_2 }$. The initial thermal distribution reads $p_2^n=\exp(-\beta_h E^n_2)/z_2$  with inverse hot temperature $\beta_h$ and partition function  $z_2$.
We have similarly for transformation (4),
\begin{equation}
	P_4(w_4)=\sum_{m}  \delta [w_4 - (E_{1}^{m}-E_{4}^{m})] p^{m}_{4},
\end{equation}  
with $p_4^m=\exp(-\beta_c E^m_4)/z_4$. In order to ensure that the system is in a thermal state at the end of each adiabat, and thus at the beginning of each isotherm, we  adjust the adiabatic driving  such that $\omega_3/  \omega_2= \omega_4/  \omega_1 ={\beta_h}/{\beta_c}$ \cite{dan20}. The whole Carnot cycle is hence quasistatic.

 Combining the contributions of all the four branches of the cycle, we find the work output distribution, 
    \begin{equation} \label{eq:pw2}
   	P(w) = \langle \delta \left[ w- \left(W - \widetilde{\Delta H_2} - \widetilde{\Delta H_4}\right)\right]\rangle,
   \end{equation}
where $W$ is given by Eq.~\eqref{1}. We have furthermore defined  the  (stochastic) difference $\widetilde{\Delta H_i}=\langle \Delta H_i \rangle - \Delta H_i$ and used the cycle condition  $\sum_i \langle \Delta H_i\rangle$ = 0. The average in Eq.~\eqref{eq:pw2} may be computed using the Boltzmann distributions at the beginning of each adiabat \cite{sup}.

\textit{Coefficient of variation for work.} In statistics, the Fano factor  (the ratio of the variance $\sigma^2$ and the mean) and the coefficient of variation (the ratio of  the standard deviation $\sigma$ and  the mean) are two  measures of the dispersion of a   probability distribution \cite{eve98}. For heat engines, the Fano factor for work, $\sigma_w^2/W$, is  equal to the quotient of the constancy  $\sigma^2_P \tau$ and the average power $P=W/\tau$ (defined over one cycle time) \cite{hol18}, while the corresponding coefficient of variation for work describes the relative work fluctuations.  All the moments of the total work   can be  evaluated from Eq.~\eqref{eq:pw2} by  integration $\langle w^n \rangle = \int dw ~ P(w) w^n$. The  variance then reads \cite{sup},
 \begin{equation}\label{7}
 	\sigma^2_w = \left(T_c - T_h\right)^2 \left[C(\beta_h,\omega_2) + C(\beta_c,\omega_4) \right],
 \end{equation}
 where we have introduced  the heat capacity of  the system, $C(\beta_j, \omega_i)=d\langle H_i\rangle/dT_j$,  at the beginning of each adiabat  \cite{sup}. We accordingly obtain the Fano factor, 
\begin{equation}\label{8}
	\frac{\sigma^2_w}{|W|} = \frac{\left(T_h - T_c\right)[C(\beta_h,\omega_2) + C(\beta_c,\omega_4)]}{\Delta S},
\end{equation}
and the corresponding coefficient of variation, 
\begin{equation}\label{9}
	\frac{\sigma_w}{|W|} = \frac{\sqrt{C(\beta_h,\omega_2) + C(\beta_c,\omega_4)}}{\Delta S}.
\end{equation}
Equations \eqref{8} and \eqref{9} describe similar physics. However,  in contrast to the Fano factor \eqref{8}, the coefficient of variation \eqref{9} has the advantage that (i) it is a dimensionless quantity that (ii)  depends solely on the heat capacities of the system  (since the entropy variation can be written as an integral of the heat capacities \cite{sup}). We shall therefore focus on that quantity in the following.

A finite-time quantum Carnot  engine with large work output and small work output fluctuations is characterized by a large inverse coefficient of variation $|W|/{\sigma_w}$. We will thus next optimize the inverse of Eq.~\eqref{9} with respect to the degree of degeneracy  and with respect to the number of levels of the working medium.
 
\begin{figure}[t]
	\centering
	\includegraphics[width=.5\textwidth]{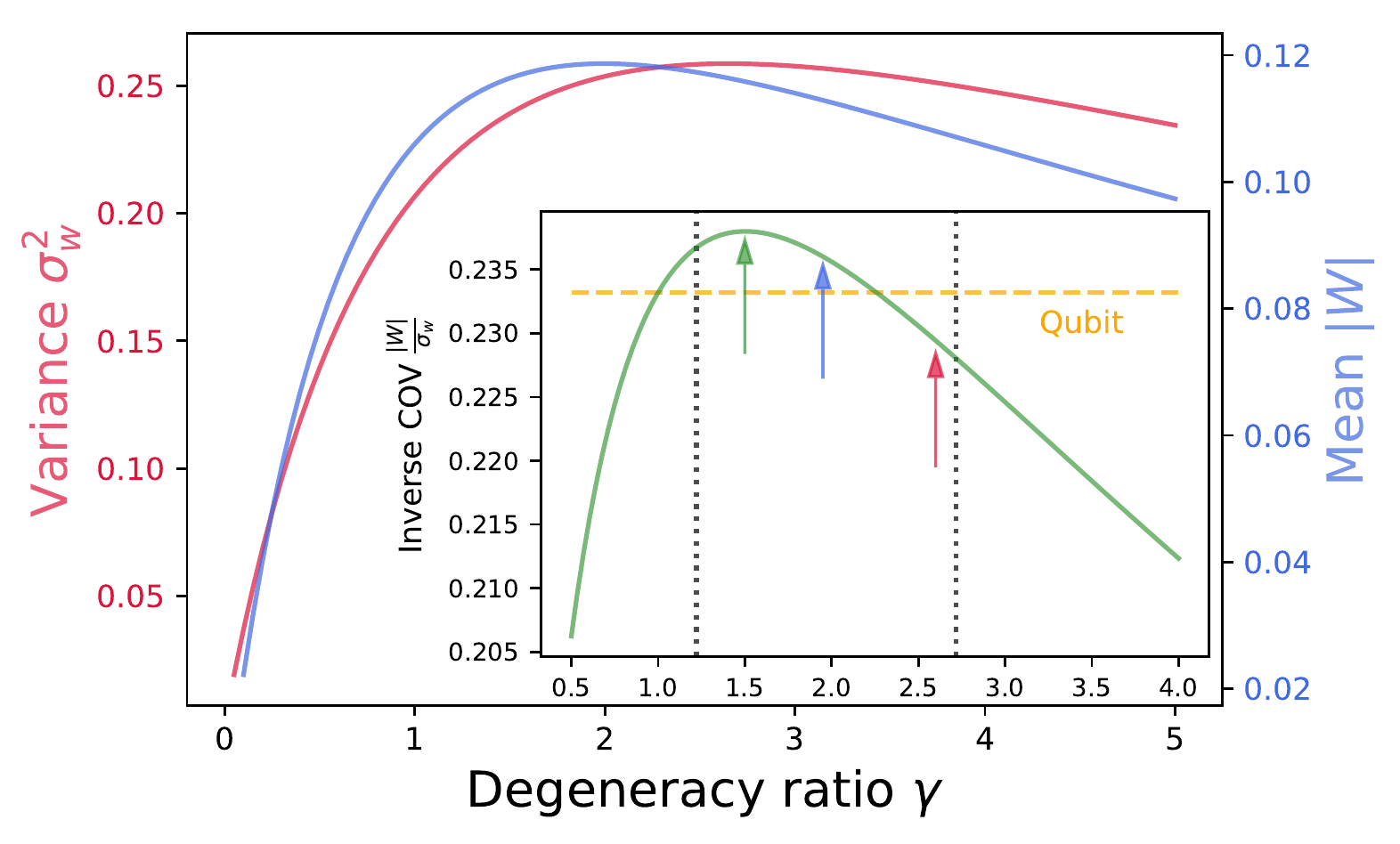}	
	\caption{Average work output $|W|$, Eq.~\eqref{1} (blue), its variance $\sigma^2_w$, Eq.~\eqref{7} (red), and  inverse coefficient of variation (COV), $|W|/\sigma_w$, Eq.~\eqref{9} (green) (inset), for a degenerate two-level quantum Carnot engine, as a function of the degeneracy ratio $\gamma$. The maximum COV (green arrow) outperforms its nondegenerate counterpart  (orange dashed line).  The blue (red) arrow indicates the maximum value of the average (variance) of the work output. The vertical black dotted lines mark  the respective  maxima of the heat capacity (Schottky anomaly), Eq. \eqref{eq:heatcap},  at the hot and cold  bath temperatures. Parameters are $\beta_c=1$, $\beta_h=0.1$, $\omega_2=2$ and $\omega_4=1$.}\label{fig:1}
\end{figure}

\textit{Degenerate two-level system.} We begin by considering  a degenerate qubit with Hamiltonian
$
H_t	=	  \omega_t g_1\ket{1}\bra{1},
$
where $g_{0}$  and $g_{1}$ are the respective degeneracies of the ground $\ket{0}$ and excited $\ket{1}$ states.  The partition function at inverse temperature $\beta$ and frequency $\omega$ is $Z_2 = g_0+ g_1 \exp(-\beta \omega)$  \cite{blu06}. The heat capacity then  follows as,
\begin{equation} \label{eq:heatcap}
	C_2(\beta,\omega) = \frac{\gamma (\beta \omega)^2}{\left(e^{\beta \omega/2} + \gamma e^{-\beta \omega/2}\right)^2},
\end{equation}
with the degeneracy ratio $\gamma=g_1/g_0$. The entropy difference during the hot isotherm further reads,
\begin{equation}
\Delta S_2=   \frac{\beta_c \omega_4 }{\gamma^{-1}e^{\beta_c \omega_4}+1}- \frac{\beta_h \omega_2 }{\gamma^{-1}e^{\beta_h \omega_2}+1} + \ln\left[\frac{1+ \gamma e^{-\beta_h \omega_2}}{1+ \gamma e^{-\beta_c \omega_4}}\right].
\end{equation}
The average of the work output  \eqref{1} (blue),  its variance \eqref{7} (red),   and the corresponding inverse coefficient of variation \eqref{9} (green) are shown  in Fig.~\ref{fig:1} as a function of the degeneracy ratio $\gamma$. We observe that,  for given frequencies and bath temperatures, both mean and  variance first increase with increasing values of $\gamma$, before they both decrease as a result of the finiteness of the Hilbert space of the qubit. However, the mean augments and decays faster than the variance. As a consequence, the inverse coefficient of variation for work exhibits a clear maximum (green arrow) for an optimal degeneracy value $\bar \gamma$. Remarkably, the degenerate quantum Carnot engine here outperforms its nondegenerate counterpart ($\gamma=1$) (orange dashed). The optimal value of the  degeneracy of the working medium may be determined by numerically solving the transcendental equation,  
\begin{eqnarray}
	&&\left(\frac{1}{2} \partial_\gamma -1\right)[C_2(\beta_c,\omega_4) + C_2(\beta_h,\omega_2)] \nonumber\\
	&& \quad+ \left(\frac{1}{2} - \partial_\gamma \right) \Delta S_2 =0.
	\end{eqnarray}
The existence of such optimal solution is guaranteed by continuity and the limiting behaviors  at $\gamma \rightarrow 0$ and $\gamma \rightarrow \infty$, where the inverse coefficient of variation vanishes.
	
Two additional features are worth emphasizing. First, the conditions of minimal relative work fluctuations and of maximum work output (blue arrow)  corresponding to,
\begin{equation}
	\bar\gamma_\text{work} = \frac{ 2 \left( e^{2x + y} - e^{x+2y} - e^{x+2y} x +e^{2x+y} y \right)}
{e^{2x}-e^{2y}+ 2 e^{x+y} \left(x-y\right) + \left(e^{x} - e^{y}\right)f(x,y)},
\end{equation}
with the variables $x=\beta_c \omega_4$ and $y=\beta_h \omega_2$, and   $f(x,y)=\sqrt{e^{2x}+e^{2y}-2e^{x+y}+4x y e^{x+y}}$,
  lead to two different solutions. While level degeneracy may be used to boost the work output \cite{all08,gel15,nie15,qua05,wan12a,wan12b}, this enhancement is accompanied by an increase of  work fluctuations, that is, of the instability of the machine.  This property might be detrimental for practical implementations of quantum heat engines. On the other hand, the point of maximum inverse coefficient of variation for work  leads to an overall smaller  work output but to a more stable engine. In addition, we note that the optimal value $\bar \gamma$ is bounded by the degeneracies associated with the respective maxima of the heat capacities (Schottky anomaly) at the hot and cold temperatures (vertical black dotted lines in Fig. ~\ref{fig:1}),
 \begin{equation}
	e^{\beta_h \omega_2} \leq \bar \gamma \leq e^{\beta_c \omega_4}.
\end{equation}
These bounds get tight when the limiting engine condition  $\omega_4 \beta_c \geq \omega_2 \beta_h $ is approached.

\textit{Nondegenerate $N$-level system.} In order to investigate the influence of the level number of the working fluid on the relative work fluctuations, we next examine  a nondegenerate $N$-level system with equidistant spacing, $H_t= \omega_t  \sum_{n=0}^{N-1} |n\rangle\langle n|$,  as appearing in homogeneous Hamiltonians of degree $-2$ \cite{gri10,jar13,cam13,def14,bea16}. The partition function at inverse temperature $\beta$ and frequency $\omega$ is given by $Z_N = [1-\exp(-N\beta \omega)]/[1-\exp(-\beta \omega)]$ \cite{blu06}. The explicit (and lengthy) expressions for the heat capacity $C_N(\beta, \omega)$ and the entropy difference $\Delta S_N$  are given in the Supplemental Material \cite{sup}. Compact expressions for the inverse coefficient of variation for work may be obtained in the limit of a harmonic oscillator ($N\rightarrow \infty$), 
\begin{equation}
\label{15}
	\left(\frac{|W|}{\sigma_w}\right)_{\infty} \!=\frac{\Delta S_{\infty}}{\sqrt{ \left(\text{sech}(y/2) y \right)^2+ \left( \text{sech}(x/2) x \right)^2}},
\end{equation}
and  for the case a (nondegenerate) qubit ($N=2$),
\begin{equation}
\label{16}
	\!\left(\frac{|W|}{\sigma_w}\right)_2\!=\frac{\Delta S_2}{\sqrt{y^2 [1-\text{tanh}(y)^2]+x^2 [1-\text{tanh}(x)^2] }}. \end{equation}
In the high-temperature limit, $\beta_{c,h} \omega_{4,2} \ll 1$, Eq.~\eqref{15} reduces to the result obtained for the classical harmonic Carnot heat engine in Ref.~\cite{hol18},\begin{equation}
	\left(\frac{|W|}{\sigma_w}\right)_{\infty}^\text{cl}=\frac{\Delta S_{\infty}}{\sqrt{2}}.
\end{equation}
On the other hand, the high-temperature limit  of Eq.~\eqref{16}  exhibits a completely different $(x,y)$-dependence, which reflects the finite Hilbert space of the qubit,
\begin{equation}
	\left(\frac{|W|}{\sigma_w}\right)_{2}^\text{cl}=\frac{\Delta S_{2}}{\sqrt{x [1-x^2]+y[1-y^2]}}.
\end{equation}
Such behavior can be traced back to the properties of the heat capacity  in Eq.~\eqref{9}: while it reaches a constant value for the (infinite-dimensional) harmonic oscillator in the classical limit (Dulong-Petit law), it vanishes for the (finite-dimensional) two-level system \cite{blu06}.

\begin{figure}[t]
	\centering
	\includegraphics[width=.5\textwidth]{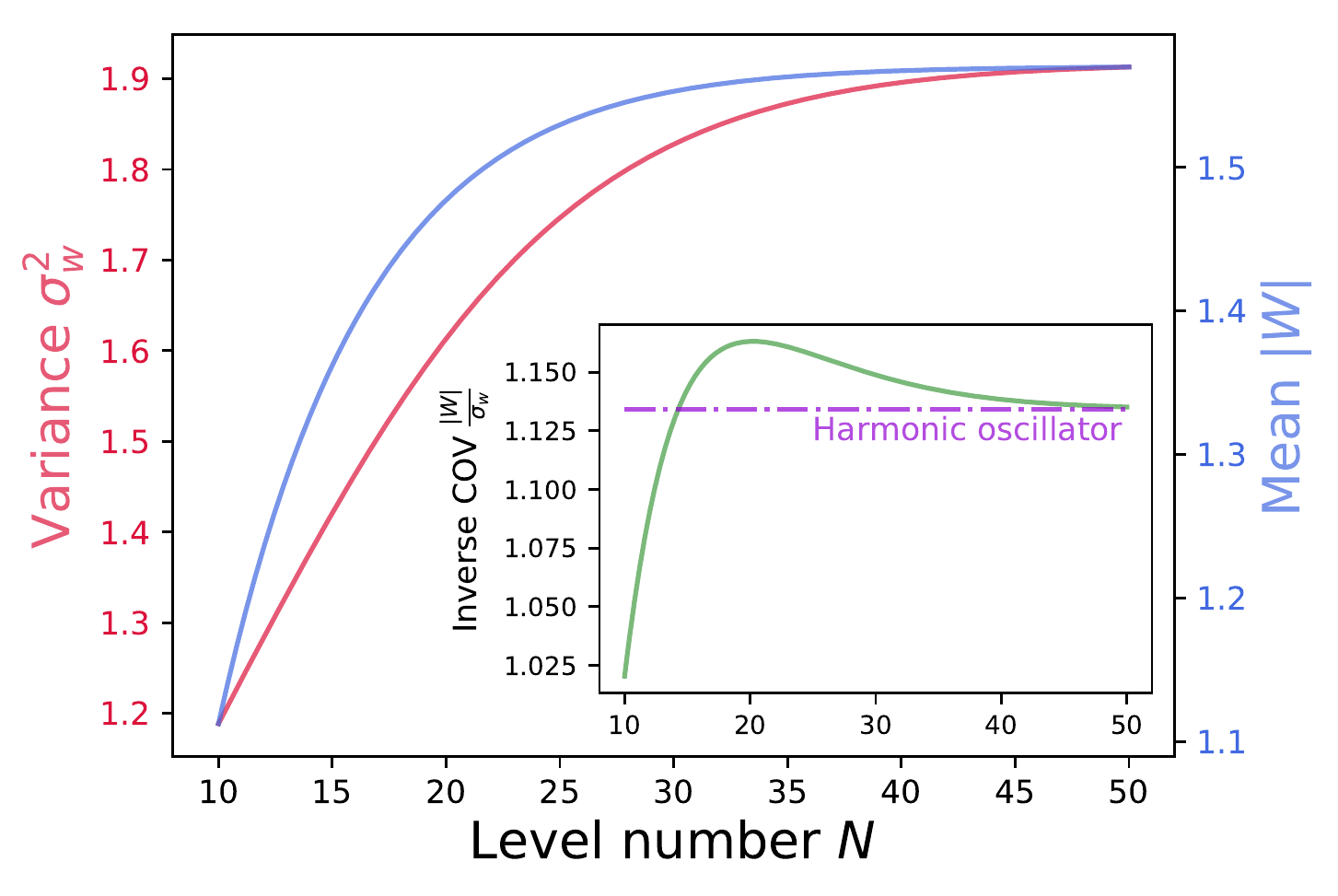}
	\caption{Average work output $|W|$, Eq.~\eqref{1} (blue), its variance $\sigma^2_w$, Eq.~\eqref{7} (red), and  inverse coefficient of variation (COV), $|W|/\sigma_w$, Eq.~\eqref{9} (green) (inset), for a nondegenerate $N$-level quantum Carnot engine, as a function of the level number $N$. The maximum COV  outperforms both that of the two-level engine and that of the harmonic oscillator motor  (violet dotted-dashed line). Same parameters as in Fig.~\ref{fig:1}.}\label{fig:2}
\end{figure}

Figure \ref{fig:2}  displays the mean work output $|W|$ (blue), the corresponding variance $\sigma^2_w$ (red)  as well as the  inverse coefficient of variation \eqref{9} (green) as a function of the level number $N$. Mean and variance increase monotonously with $N$, reaching the respective values of the harmonic oscillator in the limit $N\rightarrow \infty$. However, the mean saturates faster than the variance. The inverse relative work fluctuations therefore presents a maximum that outperforms both that of the two-level engine and of the harmonic oscillator motor.  Contrary to naive expectation, the $N$-level Carnot engine does thus not simply interpolate between these two extreme situations. The point of maximum inverse coefficient of variation  is again different from the point of maximum work output  because of increased  work fluctuations. 
The optimal level number $\bar N$ satisfies the transcendental equation,
\begin{eqnarray}
	&&\left(\frac{1}{2} \partial_N -1\right)[C_N(\beta_c,\omega_4) + C_N(\beta_h,\omega_2)] \nonumber\\
	&& \quad+ \left(\frac{1}{2} - \partial_N \right) \Delta S_N =0,
	\end{eqnarray}
which may be solved numerically.

\begin{figure}[t]
	\centering
	\includegraphics[width=.48\textwidth]{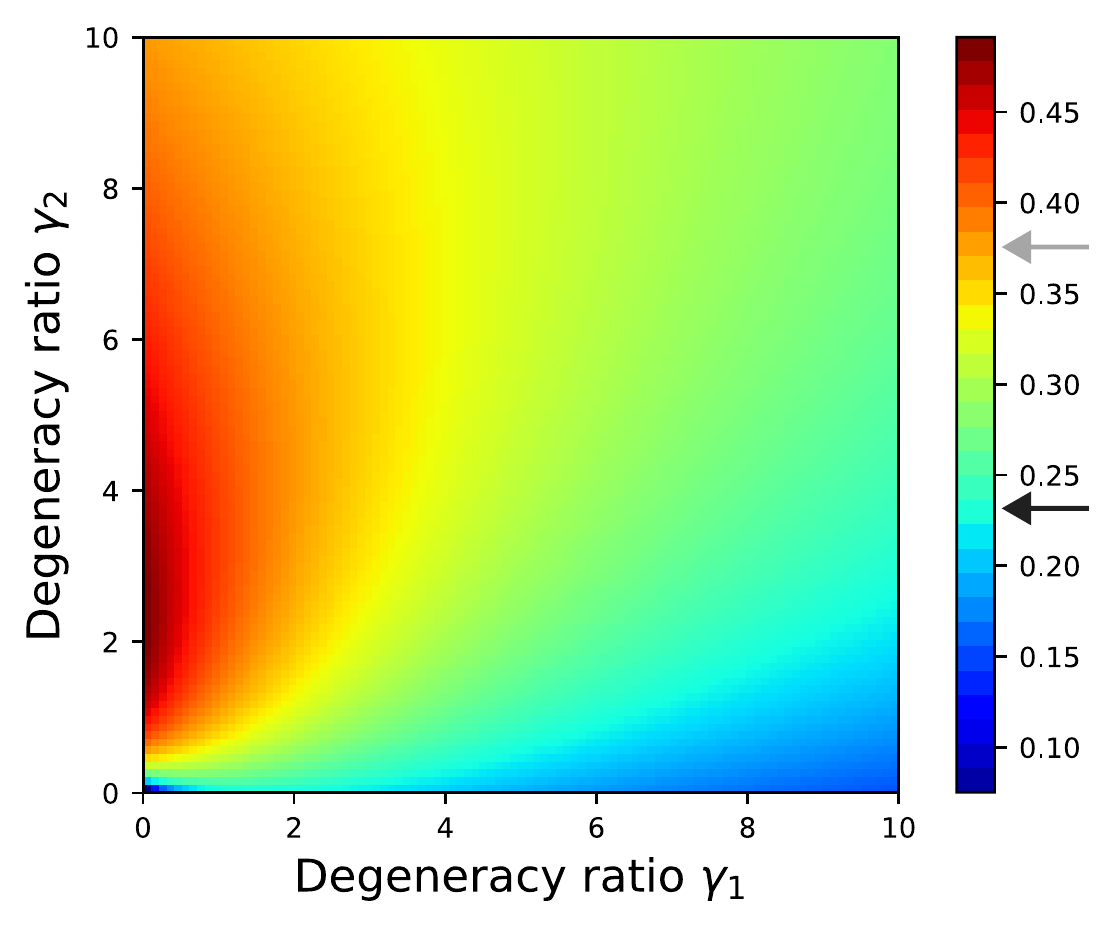}
	\caption{Inverse coefficient of variation for work, $|W|/\sigma_w$, Eq.~\eqref{9}, for a degenerate three-level quantum Carnot engine, as a function of the degeneracy ratios $\gamma_1$ and $\gamma_2$. We notice a region of high inverse coefficient of variation for small $\gamma_1$ and $1 \lesssim \gamma_2 \lesssim 4$, where nondegenerate two-level (black arrow) and three-level  heat engines (grey arrow) are outperformed. Same parameters as in Fig.~\ref{fig:1}}\label{fig:4}
\end{figure}

\textit{Degenerate $3$-level system.} We finally illustrate the usefulness of Eq.~\eqref{9} for determining the maximum inverse coefficient of variation for work for degenerate multilevel quantum Carnot engines by treating the case of a degenerate 3-level system with Hamiltonian $H_t=   \omega_t(g_1 \ket{1}\bra{1} +g_2 \ket{2}\bra{2})$ and arbitrary level degeneracies $g_n$, $(n=0, 1, 2)$. The partition function at inverse temperature $\beta$ and frequency $\omega$ is  here $Z_3 = g_0+ g_1 e^{-\beta \omega} + g_2 e^{-2\beta \omega}$. The corresponding inverse coefficient of variation for work \eqref{9} is represented as a function of  the two degeneracy ratios  $\gamma_1=g_1/g_0$ and $\gamma_2=g_2/g_0$  in Fig.~\ref{fig:4}. We identify a region of high inverse coefficient of variation for small $\gamma_1$ and $1 \lesssim \gamma_2 \lesssim 4$, where the quantum Carnot engine outperforms the respective  nondegenerate two-level (black arrow) and three-level engines (grey arrow). We moreover notice that large ratios $\gamma_1$, that is, high degeneracy of the first level, is generally detrimental to the performance of the heat engine.

\textit{Conclusions.}
Two-level systems and harmonic oscillators have been the models of choice for the investigations of  quantum heat engines in the past decades due to their simplicity \cite{gev92,gev92a,wu06,qua07,abe11,wan12,all13,dan20,abi20}. Such finite-time engines have been mostly optimized by maximizing averaged performance measures, such as mean power, with respect to cycle duration, frequency or temperature  \cite{gev92,gev92a,wu06,qua07,abe11,wan12,all13,dan20,abi20}. We have here extended these studies to include the effects of work fluctuations and of finite Hilbert space  of the working medium, two essential features of small quantum machines. To this end, we have derived a compact expression of the relative work  fluctuations of a finite-time quantum Carnot engine, as given by Eq.~\eqref{9}, in terms of the heat capacity of the system. We have shown that the  quantum motor can outperform its nondegenerate counterparts, when optimized with respect to level degeneracy or level number. We have additionally found  that optimizing the average work output, while ignoring work output fluctuations, generally leads to machines with larger instability. Our findings hence enable the analysis and future experimental realization of both high-performance and high-stability cyclic quantum heat engines. \\
\textit{Acknowledgements.} We acknowledge financial support from the Volkswagen Foundation under project "Quantum coins and nano sensors" and the German Science Foundation (DFG) under project FOR 2724.\\

\subsection*{Supplemental Material}

	\noindent \textit{Work distribution for quasistatic isotherms.} \\We show that quantum work is delta distributed, and therefore deterministic, during quasistatic isotherms, see Eq.~(3) of the main text. This result may be derived by extending the classical derivation of Ref.~\cite{hol18} to the quantum domain. We here discuss an alternative derivation by  following the train of thought presented in Ref.~\cite{sca20}. We begin by approximating the Hamiltonian evolution $H_t$ during isothermal driving as a set of $K$ discrete steps, $H_0 \rightarrow H_1 \rightarrow \dots \rightarrow H_K$. We assume the driving to be slow enough that thermal equilibrium is reached after each discrete step, so that the state of the system after the $i$-th step is of the Gibbs form, $\pi_i=e^{-\beta H_i}/Z_i$, at inverse temperature  $\beta$. The total work distribution may then be written as a convolution of $K$ independent contributions $P_i(w_i)$,
	$P(w)=\prod_{i=1}^K P_i(w_i)$.
	
	The  distribution for the $i$-th quasistatic step may be obtained via the  two-point-measurement scheme \cite{tal07} as,
	\begin{equation}\label{eq:pwi}
		P_i(w_i)=\sum_{E_{i+1}^{(m)}-E_{i}^{(n)}=w_i} \bra{E_i^{(n)}} \frac{e^{-\beta H_i}}{Z_i}\ket{E_i^{(n)}} |\braket{E_i^{(n)}| E_{i+1}^{(m)}}|^2,
	\end{equation}
	 where $\ket{E_i}$ and $E_i$ denote the respective eigenstate and  eigenenergy of the Hamiltonian $H_i$. The cumulant generating function (CGF),
	 \begin{equation}\label{eq:cgfbasic}
	 	G(\lambda) = \ln \int_{-\infty}^\infty dw ~P(w) e^{-\beta \lambda w},
	 \end{equation}
	allows the computation of all the  work cumulants via,
	\begin{equation}\label{eq:cgf_der}
		\left. \kappa_w^{l}=(-\beta)^{-l} \frac{d^l}{d\lambda^l} G(\lambda)\right |_{\lambda=0}.
	\end{equation}\\
	Inserting Eq.~\eqref{eq:pwi} into Eq.~\eqref{eq:cgfbasic}, the CGF becomes,
	\begin{eqnarray}\label{eq:cgf}
		G(\lambda) &=& \sum_{i=1}^{K-1} \int_{-\infty}^\infty dw_i~ \ln  P_i(w_i) e^{-\beta \lambda w_i} \nonumber \\
						&=&  \sum_{i=1}^{K-1} \ln \left(\text{Tr} \left[e^{-\beta \lambda H_{i+1}} e^{\beta \lambda H_i} \pi_i \right] \right) \nonumber \\
						&=& \sum_{i=1}^{K-1}  \ln \left(\text{Tr} \left[\frac{e^{-\beta \lambda H_{i+1}}}{Z_{i+1}^\lambda} \frac{e^{\beta \lambda H_i}}{Z_i^{-\lambda}} \pi_i\right] \frac{Z_{i+1}^\lambda}{Z_i^\lambda}\right) \nonumber \\
						&=& -\beta \lambda \Delta F + \sum_{i=1}^{K-1}(\lambda-1) S_\lambda \left( \pi_{i+1}|| \pi_i\right).
	\end{eqnarray}
	The CGF may thus be split into a protocol-independent part given by the free energy difference, $\Delta F=-\beta^{-1} \ln Z_K/Z_0$, and a protocol-dependent part given by the $\lambda$-Renyi divergence,
	$S_\lambda(\rho || \sigma) =  \ln \text{Tr}[\rho^\lambda \sigma^{1-\lambda}]/({\lambda - 1})$. Taking the limit $K \rightarrow \infty$ with ($H_0, H_K$)  fixed and writing the $(i+1)$-th step in terms of the $i$-th one as $H_{i+1}=H_i + \Delta H/K$, with $\Delta H=H_K-H_0$, the last term in Eq.~\eqref{eq:cgf} simplifies to,
	\begin{widetext}
	\begin{eqnarray}\label{eq:renyizero}
			\displaystyle \lim_{K \rightarrow \infty} S_\lambda \left( \pi_{i+1}||  \pi_i \right) &=& \displaystyle \lim_{K \rightarrow \infty} \frac{1}{\lambda -1}\ln \left(\frac{\text{Tr} \left[ e^{-\beta \lambda \Delta H/K} \pi_i \right]}{\text{Tr}\left[e^{-\beta (H_i + \Delta H/K)}\right]^\lambda \text{Tr}\left[e^{-\beta H_i}\right]^{-\lambda}} \right) \nonumber \\
			&=&\displaystyle \lim_{K \rightarrow \infty} \frac{1}{\lambda - 1}\left\{ \ln \left(\text{Tr} \left[ e^{-\beta \lambda \Delta H/K} \pi_i \right]\right)- \ln \left(\frac{\text{Tr}\left[e^{-\beta H_i}\right]^{\lambda}}{\text{Tr}\left[e^{-\beta (H_i + \Delta H/K)}\right]^\lambda}\right) \right\} \nonumber \\
			&=& \frac{1}{\lambda -1} \ln \text{Tr} \left[\pi_i\right] = 0,
	\end{eqnarray}	 
	\end{widetext}
	where we have used $[H_i,H_{i+1}]=0$ in the first line. All the terms hence vanish in the limit $K\rightarrow \infty$. As a result,	\begin{equation}
		G(\lambda) = -\beta \lambda \Delta F.
	\end{equation}
	The work distribution then follows as,
	\begin{equation}
		P(W) =\delta (W - \Delta F),
	\end{equation}
indicating that  work is deterministic in this case.\\

	\noindent  \textit{Derivation of the coefficient of variation for work.}\\
In order to derive Eqs.~(8) and (9) of the main text, we first need an expression for the second  moment.
Integration of Eq.~\eqref{eq:pw2}of the main text yields,
\begin{eqnarray}
	\langle w^2 \rangle &=& W^2 + \langle \Delta H_2^2 \rangle - \langle \Delta H_2 \rangle^2 +\langle \Delta H_4^2 \rangle - \langle \Delta H_4 \rangle^2. \nonumber \\
	& & 
\end{eqnarray}
So the work variance simply reads,
\begin{equation}
	\sigma_w^2 = \langle \Delta H_2^2 \rangle - \langle \Delta H_2 \rangle^2 +\langle \Delta H_4^2 \rangle - \langle \Delta H_4 \rangle^2.
\end{equation}
Using the condition, $\omega_3/  \omega_2= \omega_4/  \omega_1 ={\beta_h}/{\beta_c}$, we can simplify the mean and the square energy difference of the  adiabats in terms of the canonical partition functions,
\begin{eqnarray}
	\sigma^2_w &=& \left(1 - \frac{\beta_h}{\beta_c} \right)^2 \left[\frac{1}{z_2} \frac{\partial^2}{\partial \beta_h^2} z_2-\frac{1}{z_2^2} \left(\frac{\partial}{\partial \beta_h} z_2\right)^2 \right] \nonumber \\
	&+&\left(1 - \frac{\beta_c}{\beta_h} \right)^2 \left[\frac{1}{z_4} \frac{\partial^2}{\partial \beta_c^2} z_4-\frac{1}{z_4^2} \left(\frac{\partial}{\partial \beta_c} z_4\right)^2 \right],
\end{eqnarray}
where $z_4=\text{Tr}\left[e^{-\beta_c H_4}\right]$ and $z_2=\text{Tr}\left[e^{-\beta_h H_2}\right]$.
On the other hand, the heat capacity  at inverse temperature $\beta$ and frequency $\omega$ reads,
\begin{eqnarray}
	C(\beta, \omega) &=& \frac{\partial U(\beta, \omega)}{\partial T} = \beta^2 \frac{\partial^2}{\partial \beta^2} \ln z\\
					&=& \beta^2 \left[\frac{1}{z} \frac{\partial^2}{ \partial \beta^2}z -\frac{1}{z}\left(\frac{\partial z}{\partial \beta}\right)^2\right], \nonumber
\end{eqnarray}
Combining Eqs.~(29) and (30), we then obtain the coefficient of variation for work,
\begin{equation}\label{eq:app1}
	\frac{\sigma_w}{|W|} = \frac{\sqrt{C(\beta_c,\omega_4) + C(\beta_h,\omega_2)}}{\Delta S}.
\end{equation}
The Fano factor may be similarly  obtained by taking the square of the nominator of Eq.~\eqref{eq:app1}.

In addition, the entropy difference $\Delta S$ during the hot isotherm is given by,
 \begin{equation}\label{eq:deltas}
	\Delta S = \ln \frac{z_2}{z_4}   +T_h \frac{\partial \ln z_2}{\partial T_h} -  T_c \frac{\partial \ln z_4}{\partial T_c},
\end{equation}
or, equivalently, in terms of the heat capacities,
\begin{equation}
	\Delta S = \int_0^{T_h} dT \frac{C(1/T,\omega_2)}{T} - \int_0^{T_c} dT \frac{C(1/T,\omega_4)}{T}.
\end{equation}\\

	\noindent  \textit{Entropy variation  for harmonic oscillator and qubit.}\\
The entropy change during the hot isotherm for the harmonic oscillator ($N\rightarrow \infty$) can be determined using the corresponding partition functions $z_4$ and $z_2$,
\begin{equation}
	z_4{\Large|}_\text{HO} = \frac{1}{1- e^{-\beta_c \omega_4}} \text{ and } z_2{\Large|}_\text{HO}=\frac{1}{1- e^{-\beta_h \omega_2}}.
\end{equation}
We  find from Eq.~\eqref{eq:deltas},
\begin{eqnarray}
	\Delta S_\infty &=& -\frac{1}{2} \omega_4 \beta _c \coth \left(\frac{\omega_4 \beta _c}{2}\right)+\frac{1}{2} \beta _h \omega _2 \coth \left(\frac{1}{2} \beta _h \omega _2\right) \nonumber \\
	&-& \ln \left[\text{csch}\left(\frac{\omega_4 \beta _c}{2}\right)\right]+\ln \left[\text{csch}\left(\frac{1}{2} \beta _h \omega _2\right)\right].
\end{eqnarray}
 Similarly, the two partition functions for the two-level system ($N=2$)  are,
 \begin{equation}
 	z_4{\Large|}_\text{TLS} = 1- e^{-\beta_c \omega_4} \text{ and } z_2{\Large|}_\text{TLS}=1- e^{-\beta_h \omega_2}.
 \end{equation}
The entropy difference is accordingly,
\begin{eqnarray}
	\Delta S_{2} &=& -\omega _4 \beta _c \tanh \left(\omega _4 \beta _c\right)+\ln \left[\cosh \left(\omega _4 \beta _c\right)\right]\nonumber \\
	&+&\beta _h \omega _2 \tanh \left(\beta _h \omega _2\right)-\ln \left[\cosh \left(\beta _h \omega _2\right)\right].
\end{eqnarray}
The coefficient of variation for work for the two quantum systems can eventually be evaluated using Eq.~(31).\\
 
	\noindent  \textit{Coefficient of variation for work for a nondegenerate $N$-level system.}\\
The partition function of a $N$-level system at arbitrary temperatures and frequencies is,
\begin{equation}
	Z_{N}=\sum_{i=1}^N e^{-\beta \omega} = \frac{1-e^{-N \beta \omega}}{1-e^{-\beta \omega}}.
\end{equation}
The corresponding heat capacity reads,
\begin{widetext}
\begin{equation}
C_{N}(\beta,\omega) =	\frac{\beta^2 \omega^2 \left[e^{\beta \omega}-N^2 e^{\beta N \omega}-N^2 e^{(N+2)\beta \omega}+2 \left(N^2-1\right) e^{(N+1)\beta \omega }+e^{\beta (2 N+1) \omega}\right]}{\left(e^{\beta \omega}-1\right)^2 \left(e^{N \beta \omega}-1\right)^2}.
\end{equation}
\end{widetext}
On the other hand, the entropy difference during the hot isotherm is given by,
\begin{widetext}
	\begin{eqnarray}
		\Delta S_{N} &=& -\frac{\omega _4 \beta _c \left[N \left(-e^{\omega _4 \beta _c}\right)+e^{N \omega _4 \beta _c}+N-1\right]}{\left(e^{\omega _4 \beta _c}-1\right) \left(e^{N \omega _4 \beta _c}-1\right)}-\ln \left(\frac{1-e^{-N \omega _4 \beta _c}}{1-e^{-\omega _4 \beta_c}}\right)\nonumber \\
		&+&\frac{\beta _h \omega _2 \left[N \left(-e^{\beta _h \omega _2}\right)+e^{N \beta _h \omega _2}+N-1\right]}{\left(e^{\beta _h \omega _2}-1\right)\left(e^{N \beta _h \omega _2}-1\right)}+\ln \left(\frac{1-e^{-N \beta _h \omega _2}}{1-e^{-\beta _h \omega _2}}\right).
	\end{eqnarray}
\end{widetext}
The coefficient of variation for work again follows  by inserting  Eqs.~(39) and (40) into Eq.~\eqref{eq:app1}. We emphasize that the heat capacity and the entropy  difference exhibit a non-trivial and nonmonotonous   $N$-dependence, which is different from just increasing the number of particles for which the coefficient of variation would stay constant.

	\noindent  \textit{Coefficient of variation for work for a degenerate 3-level system.}\\
	 We finally evaluate the heat capacity and the entropy change during the hot isotherm for a degenerate 3-level system $(N=3)$.
 The partition function is,
\begin{equation}
	Z_{3} = g_0+ g_1 e^{-\beta \omega} + g_2 e^{-2\beta \omega}.
\end{equation}
The heat capacity then reads,
\begin{equation}
	C_3=(\beta \omega)^2 e^{\beta \omega}\frac{e^{2\beta \omega}\gamma_1+4 e^{\beta \omega} \gamma_2 + \gamma_1 \gamma_2}{\left(e^{2\beta\omega}+e^{\beta \omega}\gamma_1+\gamma_2 \right)^2},
\end{equation}
with the  degeneracy ratios $\gamma_1=g_1/g_0$ and $\gamma_2=g_2/g_0$. The entropy change during the hot isotherm is moreover,
\begin{widetext}
	\begin{equation}
		\Delta S_3= \frac{\beta_h \omega_2 \left(e^{\beta_h 		\omega_2}\gamma_1+2\gamma_2\right)}{e^{2 \beta_h \omega_2}+e^{\beta_h \omega_2}\gamma_1 +\gamma_2}- \frac{\beta_c \omega_4 \left(e^{\beta_c \omega_4}\gamma_1+2\gamma_2\right)}{e^{2 \beta_c \omega_4}+e^{\beta_c \omega_4}\gamma_1 +\gamma_2}+\ln\left[\frac{1+e^{-\beta_h \omega_2}(e^{\beta_h \omega_2}\gamma_1+\gamma_2)}{1+e^{-\beta_c \omega_4}(e^{\beta_c \omega_4}\gamma_1+\gamma_2)}\right].
	\end{equation}
\end{widetext}

\end{document}